\documentclass[aps,prb,twocolumn,amsmath,amssymb]{revtex4}
\usepackage{graphicx}
\usepackage{dcolumn}
\usepackage{bm}
\usepackage[usenames]{color}

\begin{document}

\title{Phonon and Electronic Non-radiative Decay of Excitons in Carbon Nanotubes}

\author{Vasili Perebeinos and Phaedon Avouris \\
IBM Research Division, T. J. Watson Research Center, \\ Yorktown
Heights, New York 10598}

\date{\today}

\begin{abstract}
We investigate theoretically the rates of non-radiative decay of
excited semiconducting nanotubes by a variety of decay mechanisms
and compare with experimental findings. We find that the
multi-phonon decay (MPD) of free excitons is too slow to be
responsible for the experimentally observed lifetimes. However, MPD
lifetimes of localized excitons could be 2-3 orders of magnitude
shorter. We also propose a new decay mechanism that relies on a
finite doping of nanotubes and involves exciton decay into an
optical phonon and an {\it intraband} electron-hole pair. The
resulting lifetime is in the range of 5 to 100 ps, even for a
moderate doping level.

\end{abstract}
\pacs{78.67.-n, 78.67.Ch,71.35.Cc} \maketitle

Semiconducting single-walled carbon nanotubes (CNTs) are
one-dimensional, direct band-gap materials with strongly-bound
exciton states, that are actively being explored for technological
applications \cite{ref1}. The potential usefulness of CNTs as an
optical material depends, to a large extent, on their luminescence
efficiency, which is determined by their radiative and nonradiative
lifetimes. The CNT radiative lifetime has been calculated
\cite{ref2,ref3} to be in the range of 1-10 ns. The measured
lifetime, however, is much shorter, in the range of 10 to 100 ps
\cite{ref4,ref5,ref6,ref7,ref8,ref9}, indicating the presence of an
efficient non-radiative decay channel(s). The nature of the dominant
non-radiative decay path has been the subject of debate
\cite{ref13,ref14,ref15}. Low energy, 'dark' exciton states
\cite{ref2,ref3,ref16,ref17} have been proposed as the cause for the
small yield, but the small dark-bright exciton splitting does not
support this conclusion \cite{ref15}. It is also known that excitons
in bulk semiconductors can be localized at structural defects,
impurities, grain boundaries and other heterogeneities \cite{ref45}.
Analogous exciton localization can occur in CNTs. In fact, the
quasi-1D, single atomic layer structure of single-walled CNTs makes
them quite sensitive to potential fluctuations in their environment,
e.g. trapped charges on an insulating substrate, adsorbed species,
etc. Recent photovoltaic studies of CNTs have directly revealed such
potential fluctuations \cite{ref20}. Even in solution, the coating
of surfactant-coated CNTs may not be homogeneous, leading to
potential fluctuations, and in otherwise perfect tubes their ends
can act as exciton traps. Photoluminescence studies of
surfactant-coated CNTs have indeed provided strong evidence for the
formation of localized excitons \cite{ref21,ref22,ref26}. The strong
effect of the environment on the non-radiative decay is most clearly
shown by the fact that no fluorescence was detected from CNTs on an
SiO$_2$ substrate \cite{ref24}, but the tubes luminescence strongly
when suspended \cite{ref25}. It is also known that, although not
intentionally doped, CNTs on most substrates in air behave as p-type
semiconductors \cite{ref14}. Suspended CNTs, on the other hand,
usually show intrinsic behavior \cite{ref27}. Furthermore, CNTs on
insulators show pronounced hysteretic effects due to shifts of their
Fermi level (field doping) by trapped charges
\cite{ref28,ref29,ref30}. Analogous charge-transfer doping effects
have been suggested for a variety of adsorbed gases on CNTs
\cite{ref31}.

It is clear that a variety of factors and environmental influences
can affect the non-radiative decay of excited nanotubes. In an
effort to evaluate their contributions, we present in this Letter a
theoretical study of the efficiency of non-radiative decay pathways
of nanotubes involving purely multi-phonon decay as well as
electronic decay mechanisms. We examine the decay behavior of
free-exciton nanotube states, trapped (localized) excitons, and
excitons in doped nanotubes, i.e. nanotubes whose Fermi level does
not coincide with the neutrality level. In the latter case, we
propose and evaluate a novel, very efficient, phonon-assisted
indirect exciton ionization (PAIEI) mechanism. Our calculations
indicate that the combination of localized exciton MPD and the PAIEI
mechanism allow us to explain the range of available experimental
data on the non-radiative lifetime. Furthermore, the calculations
suggest that, in principle, a high emission quantum yield can be
achieved in intrinsic, small diameter tubes.

\begin{figure}[h!]
\includegraphics[height=2.35in]{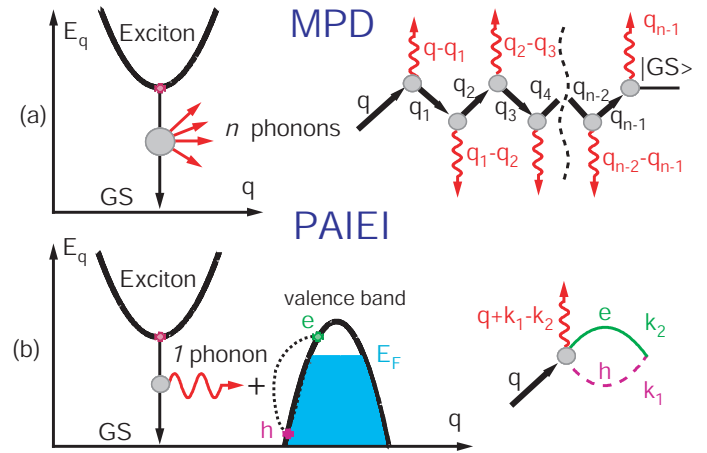}
\caption{\label{Fig1} Schematics of the exciton decay mechanisms:
(a) Multiphonon Decay (MPD) and
(b) Phonon-Assisted Indirect  Exciton Ionization (PAIEI) in p-doped CNTs.}
\end{figure}

First, we explore the MPD decay process for free excitons
schematically shown in Fig.~\ref{Fig1}a. The MPD decay of excited
CNTs to the ground state has been proposed \cite{ref13}, as being
plausible for CNTs because of the high energy C-C stretching phonon
(G-mode, $\hbar\omega\approx0.2$eV) and its strong electron-phonon
coupling \cite{ref38,ref39}.

To calculate the MPD rate we need to evaluate the $n$th order
diagram depicted in Fig.~\ref{Fig1}a \cite{ref46}. The
electron-phonon Hamiltonian involves both intraband and interband
($M^4$) electron-phonon couplings. It is obtained using the
Su-Schreiffer-Heeger model as in ref. \cite{ref38} including a
bond-bending coupling term \cite{ref40,ref46}. The exciton-phonon
coupling is derived from the electron-phonon coupling using the
two-particle exciton wavefunctions calculated from the
Bethe-Salpeter Equation as in ref. \cite{ref16,ref46}.

\begin{figure}[h!]
\includegraphics[height=3.8in]{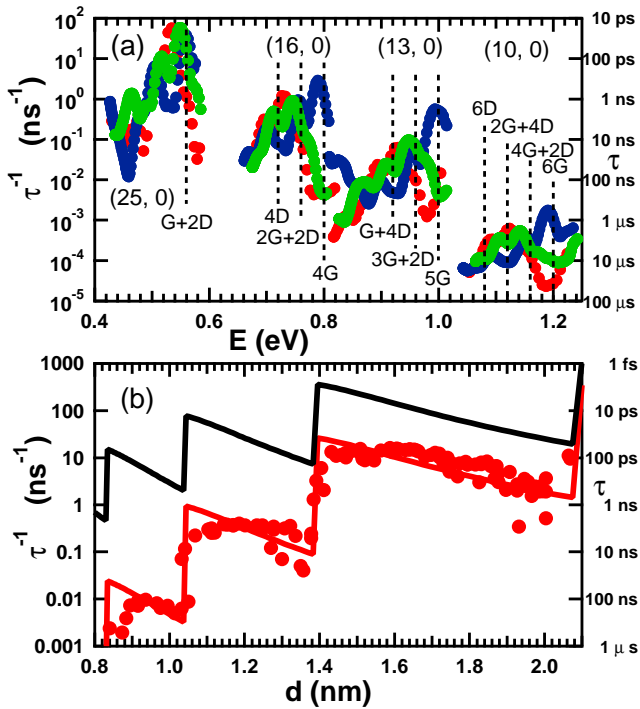}
\caption{\label{Fig2} (a) MPD rate in four nanotubes as a function of
exciton energy calculated with $\varepsilon=3.3$ for the four lowest energy bands:
bright (red), dark (blue), doubly degenerate dark (green). The vertical dashed lines
show characteristic energies of different combinations of the zone-center G mode and
even numbers of the zone-boundary D modes. (b) averaged MPD rate in
$118$ tubes as a function of tube diameter (red circles).
The solid red curve is a fit to Eq.~(\protect{\ref{eq1}}) reduced by $n!$
with $\gamma=0.04$ eV. The black curve is a simulation of the localized
exciton MPD rate.}
\end{figure}

The results of the MPD rates for the first four lowest energy {\it
free} excitons are shown in Fig.~\ref{Fig2}a. When the exciton
energy matches an integer number of optical phonon quanta, the MPD
rate shows a strong resonance, see Fig.~\ref{Fig2}a. The
Su-Schreiffer-Heeger type coupling does not break the symmetry of
the A and B carbon atom sub-lattices in the CNTs and, therefore,
does not allow coupling among even (dark) and odd (bright) parity
excitons \cite{ref2}. The bond-bending term \cite{ref40} has a much
smaller phonon coupling, but it facilitates the decay of the odd
parity states. The bright exciton ($L=0$) can decay only by emitting
phonons whose momenta sum up to zero. This is achieved by a
combination of even numbers of zone-boundary optical phonons
(D-modes of about 0.18 eV) with opposite momenta and arbitrary
numbers of Raman active G-phonons with zero momentum.

Considering the limit of a localized single exciton state coupled to a
single phonon mode, the MPD decay reduces to the classical
Franck-Condon problem, which is exactly solvable in this case  \cite{ref13}:
\begin{eqnarray}
\frac{1}{\tau_{MPD}}=\frac{2\pi}{\hbar}\frac{S^{n-1}}{\pi(n-1)!}e^{-S}\frac{n^2\gamma S_I\hbar^2\omega^2}
{\left(E-n\hbar\omega\right)^2+n^2\gamma^2}, \label{eq1}
\end{eqnarray}
where $E$ is exciton energy, $S_I$ and $S$ are the interband and
intraband Huang-Rhys factors \cite{ref41}, and $n$ is the number of
emitted phonons. The phonon density of states and phonon lifetime
are modeled by the broadening parameter $\gamma$. The interband
exciton-phonon coupling changes the number of vibrational quanta by
one, while the intraband coupling determines the overlap between
vibrationally excited state with $n-1$ phonons and vibrational
ground state.

The main difference between the calculated MPD rate and the rate
estimated according to Eq.~(\ref{eq1}) is the substantially smaller
magnitude of the former. In CNTs, in the emission process of a
phonon with momentum $q^{\prime}$  by an exciton with momentum $q$,
the phase of the exciton-phonon matrix depends on both $q^{\prime}$
and $q$. We find that cross-terms corresponding to the phonon
emissions in different orders, leading to the same final state with
$n$ phonons, will sum to nearly zero, reducing the total MPD rate by
nearly a factor of $n!$ \cite{ref46}.

We have calculated the MPD rate for {\it all chirality}
semiconducting tubes in the diameter range from 0.8 nm to 2.0 nm by
averaging the energy dependent decay rates in Fig.~\ref{Fig2}a over
the exciton distribution. The results are shown in Fig.~\ref{Fig2}b.
We find that only a few large diameter tubes, where three or less
optical phonons are needed to annihilate the exciton, have lifetimes
in the range of 100 ps to 1 ns, whereas most of the other tubes show
a much longer non-radiative lifetime. We can also fit the calculated
rates in Fig.~\ref{Fig2}b to the localized limit expression
Eq.~(\ref{eq1}) reduced by $n!$. The diameter scaling of the
intraband Huang-Rhys factor is similar to that of the optical phonon
sideband intensity \cite{ref38} $S\propto1/d$, where $d$ is a tube
diameter. The interband Huang-Rhys factor for the free e-h pair
scales inversely with the length $L$ of the tube $S_I\propto\vert
M^4\vert^2/N=A\vert M^4\vert^2/(dL)$, where we calculate $\vert
M^4\vert^2\approx0.1$ eV$^2$ and $N$ is the number of two atoms
primitive unit-cells of area $A\approx 5.39$ \AA$^2$ in the
super-cell. For a bound exciton, we find that the interband coupling
is enhanced by a factor of $L/r_0$, where the exciton radius $r_0$
depends on the diameter and effective dielectric constant
$\varepsilon$  that accounts for screening by both the nanotube
itself and the immediate environment \cite{ref16}. The choice of
$\varepsilon=3.3$ reproduces the experimentally measured exciton
binding energy in CNTs \cite{refexc}, and we find $r_0=1.95d$ and
$S_I\approx0.02/d^2$. Therefore, we are left with just one fit
parameter $S$ and the best agreement is found for $S=0.14/d$ in
Fig.~\ref{Fig2}b.

To model localized exciton MPD decay, we assume that the interband
Huang-Rhys factor $S_I$ is the same as that calculated for the free
excitons. We then calculate the intraband Huang-Rhys factor enhancement
\cite{ref41} by forcing the four lowest energy excitonic bands to
have a flat dispersion, while keeping the strength of the
exciton-phonon coupling constant. Depending on the tube diameter, we
find an effective $S$ enhancement by 1.4 to
1.8. Using Eq.~(\ref{eq1}) with $S=0.21/d$, we obtain the MPD
rate of localized excitons shown
in Fig.~\ref{Fig2}b. We find that the increased effective
exciton-phonon coupling, due to the reduced exciton dispersion, and
the coherent contributions of matrix elements bring the MPD rates
for the localized states close to the experimentally measured
non-radiative decay values, even for small diameter CNTs.

We now examine electronic non-radiative decay mechanisms. A purely
electronic (Auger) decay mechanism \cite{ref34} can be viewed as the
reverse of an impact excitation process \cite{ref35}. However, the
angular momentum conservation law sets strong restrictions for the
Auger decay, such that the E$_{11}$ bright and dark excitons with
zero angular momentum cannot decay through such a process
\cite{ref46}. The higher energy dark exciton with finite angular
momentum can decay, but only at a high doping level with the Fermi
level lying above the bottom of the second band \cite{ref46}. We
also note that at high excitation densities, exciton-exciton
annihilation dominates the decay \cite{ref34,ref36}. However, even
at low excitation densities the experimentally measured decay is
fast.

We show here that in the presence of the free carriers in CNTs, an
exciton can decay fast by creating a phonon and an intraband
electron-hole (e-h) pair, as shown in Fig.~\ref{Fig1}b. This
phonon-assisted indirect exciton ionization (PAIEI) process involves
Coulomb and exciton-phonon interactions which are both very strong
in CNTs \cite{ref16,ref32,ref33,ref38,ref39}. The interband
exciton-phonon interaction allows the exciton to cross to the
electronic ground state. However, emission of a single phonon is not
sufficient to conserve energy. The excess energy is accommodated by
producing an e-h pair in the {\it valence} band (for a p-doped CNT).
The initial excitonic state is described by a two-particle
wavefunction of exciton in state $p$ with momentum $q$: $|\Psi_q^p>
=\sum_k A^p_{kq} c^{\dagger}_{k+q}v_k |GS>$. Here, $\vert
A^p_{kq}\vert^2$ gives the contribution of the final hole state $k$
with energy $\varepsilon_k$ to the superposition of states
describing the initial exciton state. The PAIEI decay rate is given
by Fermi's golden rule as shown by the diagram in Fig.~\ref{Fig1}b
\cite{ref46} and is controlled by the overlap in reciprocal space of
the excitonic and final hole state wavefunctions $\vert
A^p_{kq}\vert^2$, and  by the Fermi level which determines the
available final states for scattering.

\begin{figure}[h!]
\includegraphics[height=2.6in]{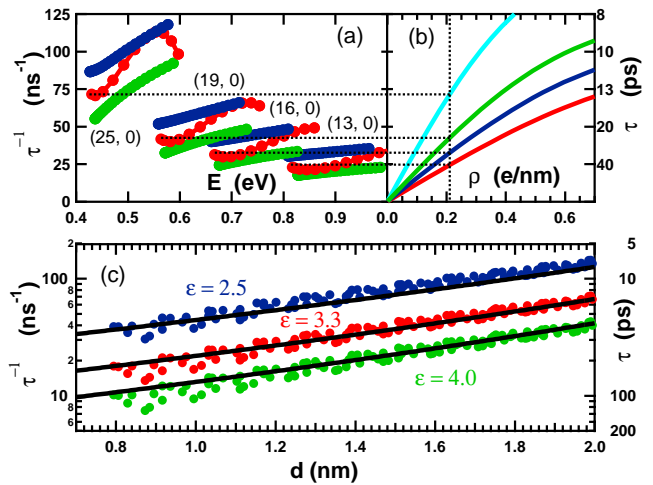}
\caption{\label{Fig3} PAIEI decay rate: (a) as a function of exciton energy
for the four lowest energy bands: bright (red), dark (blue), doubly degenerate
dark (green); (b) averaged PAIEI decay rate as a function of free carrier
density in the same tubes: red - (13, 0), blue - (16, 0), green - (19,0),
and cyan - (25, 0) using  $\varepsilon=3.3$; (c) averaged PAIEI decay rate
in $118$ tubes at constant carrier density $\rho=0.21$ e/nm,
using  $\varepsilon=2.5$ (blue circles), $\varepsilon=3.3$ (red circles),
$\varepsilon=4.0$ (green circles) along with the simplified expression
Eq.~(\protect\ref{eq2}) (solid lines).}
\end{figure}

The computed energy-dependent PAIEI rates are shown in
Fig.~\ref{Fig3}a for the four lowest energy excitons in several
nanotubes for a fixed doping level of $\rho=0.21$ e/nm. Unlike in
the MPD case, which shows resonances for an integer number of
optical phonons, the PAIEI rate shows a much weaker energy
dependence. We find that the PAIEI rate at the bottom of the each
band in {\it all chirality} tubes is largest for the lowest energy
dark exciton, while the finite angular momentum, higher energy, dark
exciton has the smallest PAIEI rate (Fig.~\ref{Fig3}a). As the
doping increases, the PAIEI rate increases proportionally
(Fig.~\ref{Fig3}b).

In the low doping limit, which is the main interest here, the EIPAI
rate can be found analytically. The number of final electron states
is proportional to the Fermi wavevector $k_F$, which defines the
free carrier density $\rho$. At zero temperature, $k_F=\pi\rho/4$.
This explains the linear dependence of the PAIEI decay rate on
$\rho$, which holds up to a 0.4 e/nm doping level. The number of
final hole states is given by the density of states, which we
calculate using hyperbolic band dispersion \cite{ref44}, at the hole
energy $\varepsilon_h\approx 3\Delta-\hbar\omega$, where
$\Delta\approx 0.42 {\rm \ eV \ nm}/d$ is half the bandgap. The tail
of the exciton wavefunction in momentum space is the key factor
determining the PAIEI rate. In the case of an uncorrelated (free)
e-h pair, the two particle wavefunction is a delta function and the
PAIEI decay cannot take place. We find that the reciprocal space
wavefunction of the lowest energy exciton is best approximated by
the following functional form $\vert A_{k,q=0}^{p=1}\vert^2\propto
(1+k^2r_0^2)^{-2.6}$ \cite{ref46}. The excitonic radius $r_0$ is a
fit parameter here and scales with tube diameter $d$ and effective
dielectric constant $\varepsilon$ as
$r_0\approx0.824d\varepsilon^{0.72}$. The analytical expression of
the PAIEI decay can then be written as:
\begin{eqnarray}
\frac{1}{\tau_{PAIEI}}=\frac{2\pi}{\hbar}\frac{AS_1\hbar^2\omega^2}{2\pi Cd}
\frac{k_hr_0}{\left(1+k_h^2r_0^2\right)^{2.6}}\frac{\rho}{\varepsilon_h}, \label{eq2}
\end{eqnarray}
where $k_h=\sqrt{\varepsilon_h^2-\Delta^2}/(\hbar V_F)$, $V_F\approx
10^6$ m/s is the Fermi velocity of graphene, $C\approx0.65$ is the
wavefunction normalization constant, $S_1$ is an effective interband
Huang-Rhys factor. Eq.~(\ref{eq2}) fits remarkably well the PAIEI
rate at room temperature and fixed carrier concentration for {\it
all chirality} semiconducting tubes with diameter range between 0.8
nm and 2.0 nm and for a variety of $\varepsilon$ with just one
adjustable parameter $S_1=2.5$ (Fig.~\ref{Fig3}c). This value of
$S_1$ suggests an electron-phonon coupling $\vert
M^4\vert^2=S_1\hbar^2\omega^2=0.1$ eV$^2$, which is in good
agreement with the calculated value here and measured in
\cite{ref30}. Finally, we also note that the PAIEI decay rate
decreases with increasing effective dielectric constant due to the
increased exciton radius.

In summary, we have calculated the non-radiative lifetimes of
excited semiconducting CNTs via multiphonon decay and by a new,
phonon-assisted electronic decay channel, as a function of the CNT
diameter.  We find that the MPD rate for free excitons is too slow,
especially for small diameter CNTs, to account for the
experimentally measured decay rates. On the other hand, we predict
that the MPD decay rate of localized excitons is increased by a few
orders of magnitude, due to a coherent contribution of
exciton-phonon matrix elements and a stronger effective
exciton-phonon coupling. Thus, localized exciton MPD rates can
account for a wide range of available experimental data. Exciton
localization is usually associated with charge transfer and doping.
In these doped tubes, we find that the decay channel involving
phonon-assisted indirect exciton ionization can dominate both the free
and localized exciton decay, even at moderately small doping levels.

\newpage

\appendix

\begin{widetext}

\section{\label{ap0} Auger decay}

The Auger decay of the exciton into electron-hole pair involves Coulomb interaction, which is
very strong in carbon nanotubes \cite{Spataru2,Perebeinos2}. In fact, Auger decay \cite{Wang} is a reverse process of the impact excitation,
which is much faster in CNTs than in bulk semiconductors \cite{PerebeinosImpact}.
However, for the Auger decay to take place both energy and momenta have to be conserved:
\begin{eqnarray}
E_{q}^{\Delta L}=e_{k+q}^{l+L}-e_{k}^{l}
\label{conserv}
\end{eqnarray}
where $E$ is the exciton energy and $e$ is the electron and hole energies. The lower index denotes the
continuous momentum along the tube axis, and the upper index the discrete angular momentum, which can have values
$l=\pm 1, \pm 2, \pm 4$ etc. The angular momentum
conservation law sets strong restrictions for the Auger decay. For example, the bright $E_{11}$ exciton with
zero angular momentum and $q=0$ can not decay by Auger. Indeed, the right hand side
of Eq.~(\ref{conserv}) is identically zero for $L=0$ and $q=0$. A similar argument applies for the lowest energy dark
exciton with $\Delta L=0$. The higher energy dark excitons \cite{PerebeinosRad} with finite angular momentum $\Delta L=2$
can decay by Auger, but in order to conserve the angular momentum, a hole has to be created in the second band $l_h=2$
and an electron in the third band $l_e=4$. This requires a high doping level, so that the Fermi level lies above the
bottom of the second band. The lowest energy exciton which can decay by Auger at low doping level is $E_{12}$ with $\Delta L=3$, which
can be photoexcited with light polarized perpendicular to the tube axis. Indeed, both momentum and energy can be
conserved when $E_{12}$ exciton decays into hole in the first band $l_h=1$ and electron in the third band
$l_e=4$ as shown in Fig.~\ref{Figap1}.

\begin{figure}[h!]
\includegraphics[height=2.35in]{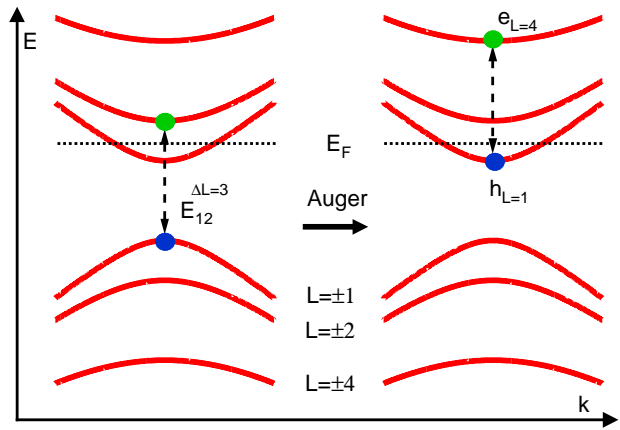}
\caption{\label{Figap1}Schematics of the Auger decay, which can take place at
low doping level for a $E_{12}$ exciton with
angular momentum  $\Delta L=3$.
The Fermi level $E_F$ is shown by the dashed line. The angular momentum conservation prevents
the optically bright exciton $E_{11}$ ($\Delta L=0$) to decay by this mechanism. The higher energy,
dark excitons with $\Delta L=2$ can decay by the Auger mechanism, but only at a high doping level with $E_F$
lying above the bottom of the second band.}
\end{figure}

\section{\label{ap4} The MPD rate calculations}

In this section we derive the multiphonon decay (MPD) rate equations using perturbation theory
expansion. The transition rate from an initial state $i$ to a final state $f$
is given as:
\begin{eqnarray}
W_{if}&=&\frac{2\pi}{\hbar}\delta\left(E_f-E_i\right)\left\vert
\left\langle f\vert T\vert i\right\rangle\right\vert^2
\nonumber \\
\left\langle f\vert T\vert i\right\rangle &=& \left\langle f\vert
V\vert i\right\rangle + \sum_k \frac{\left\langle f\vert V\vert
k\right\rangle\left\langle k\vert V\vert
i\right\rangle}{E_i-E_k+i\gamma}+ \sum_{kk'} \frac{\left\langle
f\vert V\vert k'\right\rangle\left\langle k'\vert V\vert
k\right\rangle\left\langle k\vert V\vert
i\right\rangle}{\left(E_i-E_{k'}+i\gamma\right)\left(E_i-E_k+i\gamma\right)}
+ ... , \label{trans}
\end{eqnarray}
where $V$ is the exciton-phonon potential and $\gamma$ is the level broadening.

In our case, the initial state exciton two-particle wavefunction
$\vert\Psi_q^S\rangle$ with momentum $q$ in state $S$ is given by
Eq.~(\ref{wfun}) and the final state $\vert{\rm
GS},\omega_{q_1\mu_1}, \omega_{q_2\mu_2}, \omega_{q_3\mu_3},
...\rangle$ is the electronic ground state $\vert{\rm GS}\rangle$
with a number of created phonons with all possible momenta
$q$ and band indexes $\mu$ during the decay process. The number of
created phonons in the final state determines the order of the
perturbation.

The total MPD rate of an exciton $(q, S)$ is given by the sum of the
first, second etc. orders of perturbation:
\begin{eqnarray}
W_{qS}&=&\sum_{i=1}^nW_{qS}^i \label{wtotal}
\end{eqnarray}
The lowest order non-zero contribution determines the MPD rate.
The effective decay rate is averaged over
the exciton distribution. In the case when the thermalization is
fast compared to the non-radiative lifetime,  it is given by:
\begin{eqnarray}
\bar{W}_{nr}=\frac{1}{Z(T)}\sum_{\nu}{W}_{\nu} \exp\left(
\frac{-E_{\nu}}{k_{B}T}\right), \label{wexp}
\end{eqnarray}
where index $\nu=(q,\sigma,S)$ includes summation over exciton
wavevector $q$, spin index $\sigma$ (labeling singlet and triplet
states), and exciton state $S$ (labeling the bound and continuum
states for a given $q$ and $\sigma$). The partition function is
$Z(T)=\sum_{\nu}\exp(-E_{\nu}/k_{B}T)$, summed over the same range
of $\nu$.

Below we give explicit expressions for the first, second, and third order decay rates:

First order (Fermi's Golden rule):
\begin{eqnarray}
W_{qS}^1&=&\frac{2\pi}{\hbar}\frac{1}{N}\sum_{\mu}\delta\left(E_q^S-\hbar\omega_{q\mu}\right)
\left\vert D_{q\mu}^S\right\vert^2\left(1+n_{q\mu}\right)
\label{FG1}
\end{eqnarray}

Second order:
\begin{eqnarray}
W_{qS}^2&=&\frac{2\pi}{\hbar}\frac{1}{N}\frac{1}{2!}\sum_{q_1\mu_1}\sum_{q_2\mu_2}
\delta\left(E_q^S-\hbar\omega_{q_1\mu_1}-\hbar\omega_{q_2\mu_2}\right)\delta_{q,q_1+q_2}
\left(1+n_{q_1\mu_1}\right)\left(1+n_{q_2\mu_2}\right)
\nonumber \\
&&\left\vert \sum_{S_1} \frac{B_{qq-q_1\mu_1}^{SS_1}D_{q_2\mu_2}^{S_1}}
{\left(E_q^S-E_{q-q_1}^{S_1}-\hbar\omega_{q_1\mu_1}+i\gamma\right)}+\sum_{S_2}
\frac{B_{qq-q_2\mu_2}^{SS_2}D_{q_1\mu_1}^{S_2}}
{\left(E_q^S-E_{q-q_2}^{S_2}-\hbar\omega_{q_2\mu_2}+i\gamma\right)}\right\vert^2
\label{FG2}
\end{eqnarray}

Third order:

\begin{eqnarray}
W_{qS}^3&=&\frac{2\pi}{\hbar}\frac{1}{N}\frac{1}{3!}\sum_{q_1\mu_1}\sum_{q_2\mu_2}\sum_{q_3\mu_3}
\delta\left(E_q^S-\hbar\omega_{q_1\mu_1}-\hbar\omega_{q_2\mu_2}-\hbar\omega_{q_3\mu_3}\right)\delta_{q,q_1+q_2+q_3}
\left(1+n_{q_1\mu_1}\right)\left(1+n_{q_2\mu_2}\right)\left(1+n_{q_3\mu_3}\right)
\nonumber \\
&&\left\vert P_{q_1q_2q_3}\sum_{S_1S_2} \frac{B_{qq-q_1\mu_1}^{SS_1}B_{q-q_1q-q_1-q_2\mu_2}^{S_1S_2}D_{q_3\mu_3}^{S_2}}
{\left(E_q^S-E_{q-q_1}^{S_1}-\hbar\omega_{q_1\mu_1}+i\gamma\right)
\left(E_q^S-E_{q-q_1-q_2}^{S_2}-\hbar\omega_{q_1\mu_1}-\hbar\omega_{q_2\mu_2}+i\gamma\right)}\right\vert^2
\label{FG3}
\end{eqnarray}
where $P_{q_1q_2q_3}$ stands for the sum over six permutations of indices $q_1$, $q_2$, and $q_3$. The interband
$B_{qq'\mu}^{SS'}$ and intraband $D_{q\mu}^S$ exciton phonon couplings are given in appendix \ref{ap3}.

Note, that in the absence of exciton and phonon dispersions, Eq.~(\ref{FG1})-(\ref{FG3}) reduces to the MPD rate in the
localized limit given by Eq.~(2) of the main text, except for the missing factor $e^{-S}\approx 1$, which
can not be obtained by perturbation theory.

We find that the cross terms in Eq.~(\ref{FG3}), corresponding to the different orders of
emitted phonons, sum nearly to zero. Therefore, neglecting the interference between these
terms is a good approximation to the total MPD rate:

\begin{eqnarray}
W_{qS}^n&=&\frac{2\pi}{\hbar}\sum_{q_1\mu_1}\sum_{q_2\mu_2}\ldots\sum_{\mu_n}
\delta\left(E_q^S-\hbar\omega_{q-q_1\mu_1}-\hbar\omega_{q_1-q_2\mu_2}-\ldots-\hbar\omega_{q_{n-1}\mu_{n}}\right)
\nonumber \\
&&\left(1+n_{q-q_1\mu_1}\right)\left(1+n_{q_1-q_2\mu_2}\right)\ldots\left(1+n_{q_{n-1}\mu_{n}}\right)
\left\vert {\cal K}_n\right\vert^2
\nonumber\\
{\cal K}_n&=&\sum_{S_1S_2\ldots S_{n-1}}\frac{B_{qq_1\mu_1}^{SS_1}}
{\left(E_q^S-E_{q_1}^{S_1}-\hbar\omega_{q-q_1\mu_1}+i\gamma\right)}\frac{B_{q_1q_2\mu_2}^{S_1S_2}}
{\left(E_q^S-E_{q_2}^{S_2}-\hbar\omega_{q-q_1\mu_1}-\hbar\omega_{q_1-q_2\mu_2}+i\gamma\right)}\ldots
\nonumber \\
&&\frac{B_{q_{n-2}q_{n-1}\mu_{n-1}}^{S_{n-2}S_{n-1}}} {
\left(E_q^S-E_{q_{n-1}}^{S_{n-1}}-\hbar\omega_{q-q_1\mu_1}-\hbar\omega_{q_1-q_2\mu_2}-
\ldots-\hbar\omega_{q_{n-2}-q_{n-1}\mu_{n-1}}+i\gamma\right)}
D_{q_{n-1}\mu_n}^{S_{n-1}} \label{decayn}
\end{eqnarray}

The evaluation time of Eq.~(\ref{decayn}) scales as $N^{n-1}$ which becomes computationally
prohibitive for $n\ge 4$. Alternatively, this expression is
equivalent to the following set of equations:

\begin{eqnarray}
F_1(\varepsilon q S S')&=&\frac{1}{N}\sum_{\mu}
\delta\left(\varepsilon-\hbar\omega_{q\mu}\right)
\left(1+n_{q\mu}\right)D_{q\mu}^{S}D_{q\mu}^{*S'}
\nonumber \\
F_N(\varepsilon qSS')&=&\frac{1}{N}\sum_{q_1\mu_1}\sum_{S_1S'_1}\int
d\varepsilon_1\delta\left(\varepsilon-\varepsilon_1-\hbar\omega_{q-q_1\mu_1}\right)
\left(1+n_{q-q_1\mu_1}\right)F_{N-1}(\varepsilon_1q_1S_1S'_1)
\nonumber
\\
&&\frac{B_{qq_1\mu_1}^{SS_1}B_{qq_1\mu_1}^{*S'S'_1}}
{\left(\varepsilon-\hbar\omega_{q-q_1\mu_1}-E_{q_1}^{S_1}\right)\left(\varepsilon-\hbar\omega_{q-q_1\mu_1}-E_{q_1}^{S_1'}\right)}
\nonumber \\
W_{qS}^{N}&=&\frac{2\pi}{\hbar} F_N(E_q^S,q,S,S)
\label{decayn1}
\end{eqnarray}
which are used to calculate the MPD rates in Fig. 2 and 3 of the article.

In Fig.~\ref{Figap3} we calculate the MPD rate using the exact expression,
as in Eq.~(\ref{FG1})-(\ref{FG3})
and neglecting interference terms, as in Eq.~(\ref{decayn1}). The coherent terms contribute about $1.5$ times to the
MPD rate for $n=3$ and about $2.5$ times for $n=4$, which are significantly lower than in the fully coherent case which give
a factor of $6$ larger rate for $n=3$ and a factor of $24$ for $n=4$.

In the test calculations we replaced the exciton-phonon matrix elements in Eq.~(\ref{FG3}) by their absolute values
(forcing their phases to be identically zero) and found that the MPD rate for order $n=3$ increases by a factor of $4$,
which is much closer to the fully coherent result of $3!=6$ times larger rate.

\begin{figure}[h!]
\includegraphics[height=2.35in]{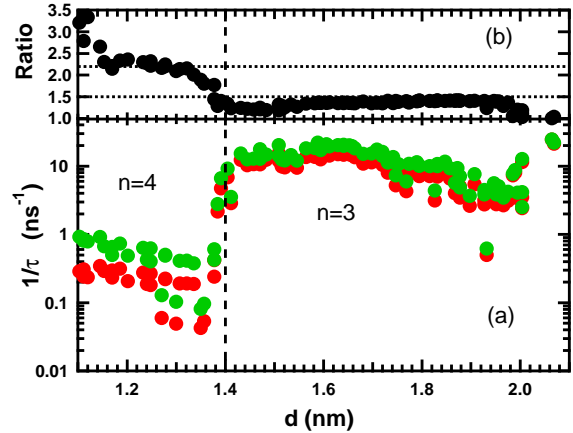}
\caption{(a) MPD decay rate of free excitons as a function of tube diameter including coherent
contributions (green circles) and neglecting them (red circles).
(b) The ratio of the two calculations (black circles) with a two dashed lines corresponding to 1.5 times and
2.2 times rate enhancement due to the coherent contributions for $n=3$ and $n=4$ correspondingly.  \label{Figap3}}
\end{figure}

\section{\label{ap5} PAIEI Decay Rate and Exciton Wavefunction }
In this section, we give the analytical form of the exciton wavefunction in the reciprocal space,
which determines the rate of the Phonon-Assisted Indirect Exciton Ionization (PAIEI). Exciton in state $p$ with momentum
$q$ and energy $E_q^p$ can PAIEI decay with the rate given by the Fermi's Golden rule:
\begin{eqnarray}
\frac{1}{\tau_{PAIEI}^{qp}}=\frac{2\pi}{N\hbar}\sum_{k_1k_2\mu}\vert M^4_{k_1+q,k_2-k_1-q,\mu}\vert^2
\vert A^p_{k_1q}\vert^2\delta\left(E_q^p-\hbar\omega_{q+k_1-k_2}-\varepsilon_{k_1}+\varepsilon_{k_2} \right)
\left(1+n_{q+k_1-k_2,\mu}\right)f_{k_1}\left(1-f_{k_2}\right)
\label{eqpaei}
\end{eqnarray}
where the Bose-Einstein factor $n_q$ gives the phonon population, while the Fermi-Dirac distribution
$f_k$ defines the probabilities that the hole state with energy and momentum $(\varepsilon_{k_1},k_1)$
and the electron state  $(\varepsilon_{k_2},k_2)$ in the {\it valence} band are available for scattering.

The exciton wavefunction $A^p_{kq}$ gives the contribution of the final hole state to the superposition
of states describing the initial exciton state. We find empirically that it is best fitted by the following equation:
\begin{eqnarray}
N\vert
A_{kq}^S\vert^2=\frac{\pi^2}{2AC}\frac{dr_0}{\left(1+(kr_0)^2\right)^{2.6}}
\label{eq5}
\end{eqnarray}
with just one adjustable parameter the exciton radius $r_0$. Here, the normalization constant
$C=\int_0^{\infty}dx/(1+x^2)^{2.6}\approx 0.65$ requires $\sum_k \vert
A_{kq}^S\vert^2=1$, and $A=0.0539$ nm$^{2}$ is the area of the two-atom primitive unit cell, $d$ is the tube diameter,
and $N$ is the number of primitive unit cells in the supercell.
The results of the fits are excellent, as shown in Fig.~\ref{Figap4} for few selected tubes. We find an empirical scaling of the
exciton radius with tube diameter and effective dielectric constant $\varepsilon$,
that accounts for screening by both the nanotube itself and the immediate environment, to be:
\begin{eqnarray}
r_0 = 0.824 \ d \ \varepsilon^{0.72}
\label{eq6}
\end{eqnarray}
which is similar to the scaling found in ref. \cite{Perebeinos2} from the real space wavefunction analysis.
\begin{figure}[h!]
\includegraphics[height=2.35in]{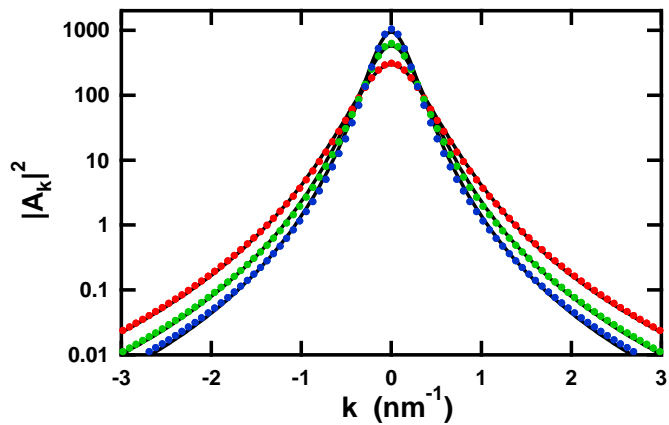}
\caption{\label{Figap4}The lowest energy excitonic wavefunctions in reciprocal space in three tubes (13,0) - red circles,
(19,0) - green circles, and (25,0) - red circles calculated for $\varepsilon=3.3$. The black solid curves
(shown behind the circles) are best fits to Eq.~(\protect{\ref{eq5}}) with $r_0$=2.05 nm, 2.71 nm, and 3.36 nm,
correspondingly.}
\end{figure}

\section{\label{ap1} SSH Electron-Phonon Matrix Elements}
In this section we give explicit form of the Su-Schreiffer-Heeger (SSH) Hamiltonian \cite{Su1,Su2} in reciprocal space.
The real space SSH Hamiltonian reads:
\begin{eqnarray}
{\cal H}_{e-ph}^{SSH}&=&g\sum_{i\delta} \delta
u_{i,i+\delta}\left(C_i^{A\dagger}C_{i+\delta}^{B}+C_{i+\delta}^{B\dagger}C_{i}^{A}
\right) \nonumber \\
C_i^{A\dagger}&=&\frac{1}{\sqrt{N}}\sum_k C_k^{A\dagger}
\exp{\left(-i\vec{k}\vec{R}_i\right)}
\nonumber \\
C_{i+\delta}^{B}&=&\frac{1}{\sqrt{N}}\sum_k C_k^{B}
\exp{\left(i\vec{k}\vec{R}_{i+\delta}\right)} \nonumber \\
\delta
u_{i,i+\delta}&=&\vec{n}_{i\delta}\left(\vec{u}_{i+\delta}^B-\vec{u}_{i}^A\right)
\nonumber \\
\vec{u}_{i}^A&=&\frac{1}{\sqrt{N}}\sum_q \vec{u}_q^{A}
\exp{\left(i\vec{q}\vec{R}_i\right)}
\nonumber \\
\vec{n}_{i\delta}&=&\vec{r}_{i+\delta}^B-\vec{r}_{i}^A,
 \label{SSHrs}
\end{eqnarray}
where $R_i$ is the position of the unit cell $i$; $r_i^A$ is the
position of atom $A$ in unit cell $i$; $N$ is the number of two
atom unit cells in the supercell; $g=5.3$ eV/\AA \ \ is the
electron-phonon coupling \cite{PerebeinosRad1}.

The SSH Hamiltonian in reciprocal space is:
\begin{eqnarray}
{\cal H}_{e-ph}^{SSH}&=&\frac{g}{\sqrt{N}}\sum_{kq\delta}
\vec{n}_{0\delta}\left(\hat{T}_{\delta}\vec{u}_q^B\exp{\left(i\vec{q}\left(\vec{R}_{\delta}-\vec{R}_0\right)\right)}-\vec{u}_q^A\right)
\nonumber \\
&&\times
\left(C_{k+q}^{A\dagger}C_{k}^{B}\exp{\left(i\vec{k}\left(\vec{R}_{\delta}-\vec{R}_0\right)\right)}+
C_{k+q}^{B\dagger}C_{k}^{A}\exp{\left(-i\left(\vec{k}+\vec{q}\right)\left(\vec{R}_{\delta}-\vec{R}_0\right)\right)}\right),
 \label{SSHqs}
\end{eqnarray}
where $\hat{T}_{\delta}$ is the rotation operator acting on displacement vector
$\vec{u}_q^B$ in the $i=0$ unit cell and rotating it to $i=\delta$ unit cell.
In the tight-binding basis:
\begin{eqnarray}
C_{k}^{A\dagger}&=&\frac{1}{\sqrt{2}}\left(v_{k}^{\dagger}+c_{k}^{\dagger}\right)
\nonumber \\
C_{k}^{B}&=&\frac{1}{\sqrt{2}}\frac{\gamma_k}{\vert\gamma_k\vert}\left(v_{k}^{}-c_{k}^{}\right)
\nonumber \\
\gamma_k&=&\sum_{\delta}\exp{\left(-i\vec{k}\left(\vec{R}_{\delta}-\vec{R}_0\right)\right)}
 \label{BareSol}
\end{eqnarray}
the electron-phonon interaction has the form:
\begin{eqnarray}
{\cal H}^{\rm SSH}_{\rm el-ph}&=&\frac{1}{\sqrt{N}}\sum_{kq\mu}\left({
M}_{kq\mu}^{1}v^{\dagger}_{k+q}v_{k}-{
M}_{kq\mu}^{2}c^{\dagger}_{k+q}c_{k}+{
M}_{kq\mu}^{3}c^{\dagger}_{k+q}v_{k}-{
M}_{kq\mu}^{4}v^{\dagger}_{k+q}c_{k}\right)
\left(a_{q\mu}+a_{-q\mu}^{\dagger}\right)
\label{sshtb}
\end{eqnarray}
where

\begin{eqnarray}
M_{kq\mu}^{1}= M_{kq\mu}^{2} &=&
\frac{g}{2}\sqrt{\frac{\hbar}{2M\omega_q^{\mu}}}\sum_{\delta}
\vec{n}_{0\delta}\left(\hat{T}_{\delta}\vec{B}_{q\mu}
\exp{\left(i\vec{q}\left(\vec{R}_{\delta}-\vec{R}_0\right)\right)}-\vec{A}_{q\mu}\right)
\nonumber \\ && \times
\left(\frac{\gamma_k}{\vert\gamma_k\vert}\exp{\left(
i\vec{k}\left(\vec{R}_{\delta}-\vec{R}_{0}\right)\right)}+
\frac{\gamma_{k+q}^*}{\vert\gamma_{k+q}\vert}\exp{\left(-
i\left(\vec{k}+\vec{q}\right)\left(\vec{R}_{\delta}-\vec{R}_{0}\right)\right)}\right)
 \nonumber \\
M_{kq\mu}^{3} = M_{kq\mu}^{4}&=&
\frac{g}{2}\sqrt{\frac{\hbar}{2M\omega_q^{\mu}}}\sum_{\delta}
\vec{n}_{0\delta}\left(\hat{T}_{\delta}\vec{B}_{q\mu}
\exp{\left(i\vec{q}\left(\vec{R}_{\delta}-\vec{R}_0\right)\right)}-\vec{A}_{q\mu}\right)
\nonumber \\ && \times
\left(\frac{\gamma_k}{\vert\gamma_k\vert}\exp{\left(
i\vec{k}\left(\vec{R}_{\delta}-\vec{R}_{0}\right)\right)}-
\frac{\gamma_{k+q}^*}{\vert\gamma_{k+q}\vert}\exp{\left(-
i\left(\vec{k}+\vec{q}\right)\left(\vec{R}_{\delta}-\vec{R}_{0}\right)\right)}\right)
 \label{SSHmatr}
\end{eqnarray}
The phonon displacements are written in the second quantized form:
\begin{eqnarray}
\vec{u}_q^A &=& \sum_{\mu}\sqrt{\frac{\hbar}{2M\omega_q^{\mu}}}
\vec{A}_{q\mu}\left(a_{q\mu}+a_{-q\mu}^{\dagger}\right) \nonumber
\\
\vec{u}_q^B &=& \sum_{\mu}\sqrt{\frac{\hbar}{2M\omega_q^{\mu}}}
\vec{B}_{q\mu}\left(a_{q\mu}+a_{-q\mu}^{\dagger}\right)
\label{secquan}
\end{eqnarray}
where $\vec{A}_{q\mu}$ and $\vec{B}_{q\mu}$ are amplitudes of the vibrations with frequency $\omega_q^{\mu}$,
and $M$ is carbon mass.

\section{\label{ap2} Bond-Bending Electron-Phonon Matrix Elements}

In this section we give explicit form of the bond-bending Hamiltonian in reciprocal space.
The bond-bending electron-phonon Hamiltonian in real space is:
\begin{eqnarray}
{\cal H}_{e-ph}^{BB}&=&g_b\sum_{i\delta_1\ne \delta_2} \Delta
\phi_A\left(\delta_1
i\delta_2\right)C_{i+\delta_1}^{B\dagger}C_{i+\delta_2}^{B}+
g_b\sum_{i\delta_1\ne \delta_2} \Delta \phi_B\left(\delta_1
i\delta_2\right)C_{i+\delta_1}^{A\dagger}C_{i+\delta_2}^{A}
 \nonumber \\
C_i^{A\dagger}&=&\frac{1}{\sqrt{N}}\sum_k C_k^{A\dagger}
\exp{\left(-i\vec{k}\vec{R}_i\right)}
\nonumber \\
 \Delta \phi_A\left(\delta_1
i\delta_2\right)&=&\frac{\cos{(\phi)}\vec{n}_{Ai\delta_1}-\vec{n}_{Ai\delta_2}}{r_{Ai\delta_1}\sin{(\phi)}}
\vec{u}^B_{i+\delta_1}+\frac{\cos{(\phi)}\vec{n}_{Ai\delta_2}-\vec{n}_{Ai\delta_1}}{r_{Ai\delta_2}\sin{(\phi)}}
\vec{u}^B_{i+\delta_2}\nonumber \\
&&+\frac{\left(r_{Ai\delta_1}-r_{Ai\delta_2}\cos{(\phi)}\right)\vec{n}_{Ai\delta_1}+
\left(r_{Ai\delta_2}-r_{Ai\delta_1}\cos{(\phi)}\right)\vec{n}_{Ai\delta_2}}{r_{Ai\delta_1}r_{Ai\delta_2}\sin{(\phi)}}
\vec{u}^A_{i}
\nonumber \\
\vec{u}_{i}^A&=&\frac{1}{\sqrt{N}}\sum_q \vec{u}_q^{A}
\exp{\left(i\vec{q}\vec{R}_i\right)}
\nonumber \\
\vec{n}_{Ai\delta}&=&\frac{\vec{r}_{Ai\delta}}{\vert
\vec{r}_{Ai\delta} \vert}
\nonumber \\
\vec{r}_{Ai\delta}&=&\vec{r}_{i+\delta}^B-\vec{r}_{i}^A, \nonumber
\\
\cos{(\phi)}&=&\vec{n}_{Ai\delta_1}\vec{n}_{Ai\delta_2}
 \label{SSHrsB}
\end{eqnarray}
where $\Delta \phi$ is the C-C-C bond angle change, $R_i$ is the position of the unit cell $i$; $r_i^A$ is the
position of atom $A$ in unit cell $i$; $N$ is the number of two atom
unit cells in the supercell; $g_b=0.9$ eV/rad  is the
electron-phonon coupling \cite{PerebeinosRad1}.

The bond-bending Hamiltonian in reciprocal space reads:
\begin{eqnarray}
{\cal
H}_{e-ph}^{Bend}&=&\frac{g_b}{\sqrt{N}}\sum_{kq\delta_1\ne\delta_2}
D_1(q\delta_1\delta_2)C_{k+q}^{B\dagger}C_{k}^{B}\exp{\left(i\vec{k}\left(\vec{R}_{\delta_2}-\vec{R}_0\right)-
i\left(\vec{k}+\vec{q}\right)\left(\vec{R}_{\delta_1}-\vec{R}_0\right)\right)}
\nonumber \\
&&+\frac{g_b}{\sqrt{N}}\sum_{kq\delta_1\ne\delta_2}
D_2(q\delta_1\delta_2)C_{k+q}^{A\dagger}C_{k}^{A}\exp{\left(i\vec{k}\left(\vec{R}_{\delta_2}-\vec{R}_0\right)-
i\left(\vec{k}+\vec{q}\right)\left(\vec{R}_{\delta_1}-\vec{R}_0\right)\right)}
\nonumber \\
D_1(q\delta_1\delta_2)&=&\frac{\cos{(\phi)}\vec{n}_{A\delta_1}-\vec{n}_{A\delta_2}}{r_{A\delta_1}\sin{(\phi)}}
\left(\hat{T}_{\delta_1}\vec{u}^B_q\right)\exp{i\vec{q}\left(\vec{R}_{\delta_1}-\vec{R}_0\right)}+
\frac{\cos{(\phi)}\vec{n}_{A\delta_2}-\vec{n}_{A\delta_1}}{r_{A\delta_2}\sin{(\phi)}}
\left(\hat{T}_{\delta_2}\vec{u}^B_q\right)\exp{i\vec{q}\left(\vec{R}_{\delta_2}-\vec{R}_0\right)}\nonumber \\
&&+\frac{\left(r_{A\delta_1}-r_{A\delta_2}\cos{(\phi)}\right)\vec{n}_{A\delta_1}+
\left(r_{A\delta_2}-r_{A\delta_1}\cos{(\phi)}\right)\vec{n}_{A\delta_2}}{r_{A\delta_1}r_{A\delta_2}\sin{(\phi)}}
\vec{u}^A_{q}
\nonumber \\
D_2(q\delta_1\delta_2)&=&\frac{\cos{(\phi)}\vec{n}_{B\delta_1}-\vec{n}_{B\delta_2}}{r_{B\delta_1}\sin{(\phi)}}
\left(\hat{T}_{\delta_1}\vec{u}^A_q\right)\exp{i\vec{q}\left(\vec{R}_{\delta_1}-\vec{R}_0\right)}+
\frac{\cos{(\phi)}\vec{n}_{B\delta_2}-\vec{n}_{B\delta_1}}{r_{B\delta_2}\sin{(\phi)}}
\left(\hat{T}_{\delta_2}\vec{u}^A_q\right)\exp{i\vec{q}\left(\vec{R}_{\delta_2}-\vec{R}_0\right)}\nonumber \\
&&+\frac{\left(r_{B\delta_1}-r_{B\delta_2}\cos{(\phi)}\right)\vec{n}_{B\delta_1}+
\left(r_{B\delta_2}-r_{B\delta_1}\cos{(\phi)}\right)\vec{n}_{B\delta_2}}{r_{B\delta_1}r_{B\delta_2}\sin{(\phi)}}
\vec{u}^B_{q}
 \label{SSHqsB}
\end{eqnarray}
where $\hat{T}_{\delta}$ is the rotation operator, which rotates carbon atom in the original
unit cell to the carbon atom in the neighboring unit cell $\delta$. In the tight-binding basis Eq.~(\ref{BareSol})
the bond-bending coupling is:
\begin{eqnarray}
{\cal H}_{\rm el-ph}=\frac{1}{\sqrt{N}}\sum_{kq\mu}\left({
M}_{kq\mu}^{1}v^{\dagger}_{k+q}v_{k}-{
M}_{kq\mu}^{2}c^{\dagger}_{k+q}c_{k}+{
M}_{kq\mu}^{3}c^{\dagger}_{k+q}v_{k}-{
M}_{kq\mu}^{4}v^{\dagger}_{k+q}c_{k}\right)
\left(a_{q\mu}+a_{-q\mu}^{\dagger}\right), \label{HephB}
\end{eqnarray}
 where
\begin{eqnarray}
M_{kq\mu}^{1}= -M_{kq\mu}^{2} &=&F_2(kq\mu)
+F_1(kq\mu)\frac{\gamma_k}{\vert\gamma_k\vert}\frac{\gamma^*_{k+q}}{\vert\gamma_{k+q}\vert}
 \nonumber \\
M_{kq\mu}^{3} = -M_{kq\mu}^{4}&=&F_2(kq\mu)
-F_1(kq\mu)\frac{\gamma_k}{\vert\gamma_k\vert}\frac{\gamma^*_{k+q}}{\vert\gamma_{k+q}\vert}
 \label{SSHmatrB}
\end{eqnarray}
where
\begin{eqnarray}
F_1(kq\mu)&=&\frac{g_b}{2}\sqrt{\frac{\hbar}{2M\omega_q^{\mu}}}\sum_{\delta_1\ne\delta_2}\exp{\left(i\vec{k}\left(\vec{R}_{\delta_2}-\vec{R}_0\right)-
i\left(\vec{k}+\vec{q}\right)\left(\vec{R}_{\delta_1}-\vec{R}_0\right)\right)}
\nonumber \\
&&\times
[\frac{\cos{(\phi)}\vec{n}_{A\delta_1}-\vec{n}_{A\delta_2}}{r_{A\delta_1}\sin{(\phi)}}
\left(\hat{T}_{\delta_1}\vec{B}_{q\mu}\right)\exp{i\vec{q}\left(\vec{R}_{\delta_1}-\vec{R}_0\right)}
\nonumber \\
&&+\frac{\cos{(\phi)}\vec{n}_{A\delta_2}-\vec{n}_{A\delta_1}}{r_{A\delta_2}\sin{(\phi)}}
\left(\hat{T}_{\delta_2}\vec{B}_{q\mu}\right)\exp{i\vec{q}\left(\vec{R}_{\delta_2}-\vec{R}_0\right)}
\nonumber \\
&&+\frac{\left(r_{A\delta_1}-r_{A\delta_2}\cos{(\phi)}\right)\vec{n}_{A\delta_1}+
\left(r_{A\delta_2}-r_{A\delta_1}\cos{(\phi)}\right)\vec{n}_{A\delta_2}}{r_{A\delta_1}r_{A\delta_2}\sin{(\phi)}}
\vec{A}_{q\mu}]\nonumber \\
F_2(kq\mu)&=&\frac{g_b}{2}\sqrt{\frac{\hbar}{2M\omega_q^{\mu}}}\sum_{\delta_1\ne\delta_2}\exp{\left(i\vec{k}\left(\vec{R}_{\delta_2}-\vec{R}_0\right)-
i\left(\vec{k}+\vec{q}\right)\left(\vec{R}_{\delta_1}-\vec{R}_0\right)\right)}
\nonumber \\
&&\times
[\frac{\cos{(\phi)}\vec{n}_{B\delta_1}-\vec{n}_{B\delta_2}}{r_{B\delta_1}\sin{(\phi)}}
\left(\hat{T}_{\delta_1}\vec{A}_{q\mu}\right)\exp{i\vec{q}\left(\vec{R}_{\delta_1}-\vec{R}_0\right)}
\nonumber \\
&&+
\frac{\cos{(\phi)}\vec{n}_{B\delta_2}-\vec{n}_{B\delta_1}}{r_{B\delta_2}\sin{(\phi)}}
\left(\hat{T}_{\delta_2}\vec{A}_{q\mu}\right)\exp{i\vec{q}\left(\vec{R}_{\delta_2}-\vec{R}_0\right)}\nonumber \\
&&+\frac{\left(r_{B\delta_1}-r_{B\delta_2}\cos{(\phi)}\right)\vec{n}_{B\delta_1}+
\left(r_{B\delta_2}-r_{B\delta_1}\cos{(\phi)}\right)\vec{n}_{B\delta_2}}{r_{B\delta_1}r_{B\delta_2}\sin{(\phi)}}
\vec{B}_{q\mu}]
 \label{SSHmatrB1}
\end{eqnarray}

\section{\label{ap3} Exciton-Phonon Matrix Elements}
In this section we derive the exciton-phonon Hamiltonian starting with
the electron-phonon Hamiltonian:
\begin{eqnarray}
{\cal H}_{\rm el-ph}=\frac{1}{\sqrt{N}}\sum_{kq\mu}\left({
M}_{kq\mu}^{1}v^{\dagger}_{k+q}v_{k}-{
M}_{kq\mu}^{2}c^{\dagger}_{k+q}c_{k}+{
M}_{kq\mu}^{3}c^{\dagger}_{k+q}v_{k}-{
M}_{kq\mu}^{4}v^{\dagger}_{k+q}c_{k}\right)
\left(a_{q\mu}+a_{-q\mu}^{\dagger}\right), \label{Heph}
\end{eqnarray}
where ${ M}_{kq\mu}^{i}\propto g$ is momentum dependent
electron-phonon coupling (see appendixes \ref{ap1} and \ref{ap2}); $a^{\dagger}_{-q\mu}$ is a phonon
creation operator with wavevector $-q$ and phonon band index
$\mu=1...6$; and $N$ is the number of primitive unit cells, each
containing two carbons; $c^{\dagger}_{k+q}$ ($v_{k}$) creation
(annihilation) of an electron in the conduction (valence) band
acting on the ground state $\left\vert{\rm
GS}\right\rangle=\prod_k v^{\dagger}_k\left\vert{\rm
vac}\right\rangle$. The indices $k$ and $q$ label both the
continuous 1D wavevector along the tube axis and the band index.

To derive the exciton-phonon matrix elements from Eq.~(\ref{Heph}) we
need to know the exciton wavefunction given by the solution of the
Bethe-Salpeter Equation (BSE) for the two-particle exciton wavefunction
\cite{Spataru2,Perebeinos2}:
\begin{eqnarray}
\left\vert\Psi_{q}^S\right\rangle=\sum_{k}
A^S_{kq}c^{\dagger}_{k+q}v_{k}\left\vert{\rm GS}\right\rangle.
\label{wfun}
\end{eqnarray}
Here $A^S_{kq}$ is the eigenvector of the $S$'s state of BSE
solution. Using the orthogonality relation of the BSE solution
$(\sum_S A_{kq}^SA_{k'q}^{S*}=\delta_{kk'})$,
we obtain:
\begin{eqnarray}
c^{\dagger}_{k+q}v_{k}=\sum_S
A_{kq}^{S*}\left\vert\Psi_{q}^S\right\rangle. \label{free}
\end{eqnarray}

Using Eq.~(\ref{Heph}-\ref{free}), we derive the exciton-phonon
Hamiltonian:
\begin{eqnarray}
{\cal H}_{\rm
el-ph}\left\vert\Psi_q^S\right\rangle&=&\frac{1}{\sqrt{N}}\sum_{S'q'\mu}
B^{SS'}_{qq'\mu} \left(\sqrt{n_{q-q'\mu}+1}\left\vert
n_{q-q'\mu}+1\right\rangle+\sqrt{n_{q'-q\mu}}\left\vert
n_{q'-q\mu}-1\right\rangle\right)\left\vert\Psi_{q'}^{S'}\right\rangle \nonumber\\
&&+\frac{1}{\sqrt{N}}\sum_{\mu}D^{S}_{q\mu}
\left(\sqrt{n_{q\mu}+1}\left\vert
n_{q\mu}+1\right\rangle+\sqrt{n_{-q\mu}}\left\vert
n_{-q\mu}-1\right\rangle\right)\left\vert {\rm GS}\right\rangle
\nonumber \\
B^{SS'}_{qq'\mu}&=&-\sum_k \left[{
M}^{1}_{kq'-q\mu}A_{k+q'-qq}^{S}A_{kq'}^{S'*} + {
M}^{2}_{k+qq'-q\mu}A_{k,q}^{S}A_{kq'}^{S'*}\right]
\nonumber\\
D^{S}_{q\mu}&=&-\sum_k { M}^{4}_{k+q,-q\mu}A_{kq}^{S},
\label{excph}
\end{eqnarray}
where $n_{q\mu}$ is the Bose-Einstein phonon occupation number.
Here we neglect the coupling of the one exciton subspace to the subspace
with two excitons. Matrix elements $B^{SS'}_{qq'\mu}$ are the intraband and
$D^{S}_{q\mu}$ are the interband exciton-phonon couplings.

\end{widetext}


\begin{thebibliography}{}


\bibitem{ref1} Ph. Avouris, Z. Chen, V. Perebeinos, Nature Nano {\bf 2}, 605 (2007).

\bibitem{ref2} V. Perebeinos, J. Tersoff, and Ph. Avouris, Nano Lett. {\bf 5}, 2495 (2005).

\bibitem{ref3}  C. D. Spataru et al., Phys. Rev. Lett. {\bf 95}, 247402 (2005).

\bibitem{ref8} A. Hagen et al., Appl. Phys. A {\bf 78}, 1137 (2004).

\bibitem{ref4} Y.-Z. Ma et al., J. Chem. Phys. {\bf 120}, 3368 (2004).

\bibitem{ref5} G. N. Ostojic et al., Phys. Rev. Lett. {\bf 92}, 117402 (2004).

\bibitem{ref6} L. Huang, H. N. Pedrosa, and T. D. Krauss, Phys. Rev. Lett. {\bf 93}, 017403 (2004).

\bibitem{ref7} F. Wang et al., Phys. Rev. Lett. {\bf 92}, 177401 (2004).

\bibitem{ref9} M. Jones et al., Phys. Rev. B {\bf 71}, 115426 (2005).




\bibitem{ref13} Ph. Avouris et al.  Physica Status Solidi B {\bf 243}, 3197 (2006).

\bibitem{ref14} {\it Carbon Nanotubes, Topics in Applied Physics}, edited by A. Jorio, M.S.
Dresselhaus, and G. Dresselhaus  (Springer-Verlag, New York, 2007).

\bibitem{ref15} J. Shaver and J. Kono, Laser \& Photonics Reviews {\bf 1}, 260 (2007).


\bibitem{ref16} V. Perebeinos, J. Tersoff, and Ph. Avouris, Phys. Rev. Lett. {\bf 92}, 257402 (2004).

\bibitem{ref17} H. Zhao, and S. Mazumdar, Phys. Rev. Lett. {\bf 93}, 157402 (2004).

\bibitem{ref45} P.Y. Yu and M. Cardona in {\it Fundamentals of
Semiconductors} (Springer-Verlag, Berlin, 1999).




\bibitem{ref20} M. Freitag et al., Appl. Phys. Lett. {\bf 91}, 031101 (2007).

\bibitem{ref21} L. Cognet et al., Science {\bf 316}, 1465 (2007).

\bibitem{ref22} D. A. Tsyboulski et al., Nano Lett. {\bf 7}, 3080 (2007).

\bibitem{ref26} A. Hagen et al., Phys. Rev. Lett. {\bf 95}, 197401 (2005).



\bibitem{ref24} J. Lefebvre, Y. Homma, and P. Finnie, Phys. Rev. Lett. {\bf 90}, 217401 (2003).

\bibitem{ref25} J. Lefebvre et al., Appl. Phys. A {\bf 78}, 1107 (2004).



\bibitem{ref27} E. D. Minot et al., Nature {\bf 428}, 536 (2004).

\bibitem{ref28} M. S. Fuhrer et al., Nano Lett. {\bf 2}, 755 (2002).

\bibitem{ref29} W. Kim et al., Nano Lett. {\bf 3}, 193 (2003).

\bibitem{ref30} J. C. Tsang et al., Nature Nano {\bf 2}, 725 (2007).

\bibitem{ref31} R. J. Chen et al., PNAS {\bf 100}, 4984 (2003).

\bibitem{ref38} V. Perebeinos, J. Tersoff, and Ph. Avouris, Phys. Rev. Lett {\bf 94}, 027402 (2005).

\bibitem{ref39} M. Lazzeri et al., Phys. Rev. Lett. {\bf 95}, 236802 (2005).

\bibitem{ref46} see EPAPS Document No. (appendix attached)

\bibitem{ref40} V. Perebeinos, P. B. Allen, and M. Pederson, Phys. Rev. A {\bf 72}, 012501 (2005).

\bibitem{ref41} M. Ueta et al., in {\it Excitonic Processes in Solids} (Springer-Verlag, Berlin, 2000), Vol. 60.

\bibitem{refexc} G. Dukovic et al., Nano Lett. {\bf 5}, 2314 (2005).

\bibitem{ref34} F. Wang et al., Phys. Rev. B {\bf 70}, 241403 (R) (2004).

\bibitem{ref35} V. Perebeinos, and Ph. Avouris, Phys. Rev. B {\bf 74}, 121410 (R) (2006).

\bibitem{ref36} Y.-Z. Ma et al., Phys. Rev. Lett. {\bf 94}, 157402 (2005).

\bibitem{ref32}    T. J. Ando, Phys. Soc. Jpn. {\bf 66}, 1066 (1997).

\bibitem{ref33} C. D. Spataru et al., Phys. Rev. Lett {\bf 92}, 077402 (2004).













\bibitem{ref44} J. W. Mintmire, B. I. Dunlap, and C. T. White, Phys. Rev. Lett. {\bf 68}, 631 (1992).




\end{thebibliography}

\begin{thebibliography}{}

\bibitem{Spataru2} Spataru, C. D., Ismail-Beigi, S., Benedict, L. X., and Louie, S. G.
Excitonic effects and optical spectra of single-walled carbon nanotubes.
{\it Phys. Rev. Lett} {\bf 92}, 077402 (2004).
\bibitem{Perebeinos2} Perebeinos, V., Tersoff, J., and Avouris, P.
Scaling of excitons in carbon nanotubes. {\it Phys. Rev. Lett.} {\bf 92}, 257402 (2004).
\bibitem{Wang} Wang, F., Dukovic, G., Knoesel, E., Brus, L. E., and Heinz, T. F.
Observation of rapid Auger recombination in optically excited semiconducting carbon nanotubes.
{\it Phys. Rev. B} {\bf 70}, 241403 (2004).
\bibitem{PerebeinosImpact} Perebeinos, V. and Avouris, P. Impact excitation by hot carriers in carbon nanotubes.
{\it Phys. Rev. B} {\bf 74}, 121410 (2006).

\bibitem{PerebeinosRad} Perebeinos, V., Tersoff, J., and Avouris, P.
Radiative lifetime of excitons in carbon nanotubes. {\it Nano Lett.} {\bf  5}, 2495-2499 (2005).

\bibitem{Su1} Su, W. P., Schrieffer, J. R., and Heeger, A. J.
Solitons in Polyacetylene. {\it  Phys. Rev. Lett.} {\bf 42}, 1698-1701 (1979).
\bibitem{Su2} Su, W. P., Schrieffer, J. R., and Heeger, A. J.
Soliton excitations in polyacetylene. {\it  Phys. Rev. B} {\bf 22}, 2099 - 2111 (1980).

\bibitem{PerebeinosRad1} Perebeinos, V., Allen, P. B., and Pederson, M.
Reexamination of the Jahn-Teller instability in C$_6$H$_6^+$ and C$_6$H$_6^-$.
{\it Physical Review A} {\bf 72}, 012501 (2005).

\end{thebibliography}
\end{document}